# Active nano-mechanical stimulation of single cells for mechanobiology

M. Monticelli[1], D. S. Jokhun[2], D. Petti[1], G. V. Shivashankar[2,3,4] and R. Bertacco[1,5]

[1]Department of Physics, Politecnico di Milano, Milan, Italy
[2]Mechanobiology Institute, National University of Singapore, Singapore
[3]Department of Biological Sciences, National University of Singapore, Singapore
[4]Institute of Molecular Oncology, Italian Foundation for Cancer Research, Milan, Italy
[5]IFN-CNR, c/o Politecnico di Milano, Milan, Italy

**Abstract**

In-vivo, cells are frequently exposed to multiple mechanical stimuli arising from the extracellular microenvironment, with deep impact on many biological functions. On the other hand, current methods for *mechanobiology* do not allow to easily replicate in-vitro the complex spatio-temporal behavior of such mechanical signals. Here, we introduce a new platform for studying the mechanical coupling between the extracellular environment and the nucleus in living cells, based on active substrates for cell culture made of *Fe*-coated polymeric micropillars. Under the action of quasi-static external magnetic fields, each group of pillars produces synchronous nano-mechanical stimuli at different points of the cell membrane, thanks to the highly controllable pillars' deflection. This method enables to perform a new set of experiments for the investigation of cellular dynamics and mechanotransduction mechanisms in response to a periodic *nano-pinching*, with tunable strength and arbitrary temporal shape. We applied this methodology to NIH3T3 cells, demonstrating how the nano-mechanical stimulation affects the actin cytoskeleton, nuclear morphology, H2B core-histone dynamics and MKL transcription-cofactor nuclear to cytoplasmic translocation.

**Introduction**

In the last two decades, a growing scientific interest has been attracted by the emerging field of *mechanobiology*, which aims at studying the modifications of cell properties, and related transduction mechanisms, occurring when cells sense and respond to mechanical stimuli. Recent works[1,2,3,4] have highlighted how infected and mutated cells exhibit altered mechanical properties and specific mechanically activated biochemical pathways, whose understanding can be crucial for the diagnosis and treatment of several diseases (e.g. cancer).

In particular, the mechanical interactions between extracellular matrix and cells play a fundamental role in regulating cells behaviors such as migration[5], differentiation[6] and proliferation[7]. In these cellular processes, matrix signals are transduced to the nucleus eventually resulting in alterations of gene expression[8]. Various studies aim at investigating the nuclear response to mechanical stimuli applied on the cell membrane, and explore how the cytoskeleton mediates mechanical force transduction from the peripheral area to the nucleus[9,10,11]. However, a precise understanding of these mechanisms is still limited by the inherent difficulties to reproduce in-vitro the stress fields applied to the cell in-vivo. The most common methods for studying cell mechanotransduction include atomic force microscopy[12], magnetic[13,14,15,16] and optical tweezers[17], micropipette aspiration[18,19], parallel plate compression[20] and induced uniform strain on deformable substrates[21]. A few studies have shown that cell nuclei change shape in response to physical confinement, like that induced by microposts[22,23,24], micropatterned adhesive molecules[25,26] and constrictions in microfluidic channels[27]. Polymeric substrates with active functionalities[28,29] have been proposed, also including magnetic microposts[30]. However, the aforementioned techniques are unable to simultaneously apply controlled and localized forces at multiple points on the cell, with tunable spatio-temporal behavior suitable to reproduce in-vitro mechanical stimuli from the extracellular matrix[31].

Here we present a novel platform for mechanobiology: an active substrate for cell culture consisting in an array of groups of PDMS pillars with magnetic heads. Under the action of an external quasi-static magnetic field, all groups of pillars stretch and retract synchronously, thus exerting a sort of local "nano-pinching" on the cell membrane, at the points of focal adhesion. In the specific case of square groups of *Fe*-coated micropillars, a rotating magnetic field in the plane of the substrate produces a continuous biaxial deformation of the pillars, up to a maximum deflection of 600 nm for an applied field of 100 mT. This produces a periodic and local cell pinching with a maximum strain of 5% on the cell membrane. By tuning both the magnitude and direction of the external magnetic field, our platform allows for the real-time control of the intensity and temporal behavior of the applied stress field.

The potential of this method has been assessed in experiments aiming at studying the nuclear dynamic response to the periodic nano-pinching produced by magnetic pillars. The prolonged application of periodic local forces on single fibroblast cells, with amplitude of a few tens of nN and frequency of 0.1 Hz, affects the nuclear morphology and deformability, as well as the turnover of H2B core-histone, a protein of the chromatin. In addition, the nucleus-pillars coupling is not purely mechanic but mediated by active cellular mechanisms involving cytoskeleton reorganization and biomolecular processes across the nuclear membrane. This emerges from the observed enhancement in actin dynamics and translocation of MKL transcription cofactor from the nucleus to the cytoplasm during stimulation.

Our results show that the magnetic nano-pinching provides a novel approach to the in-vitro study of cellular mechanotransduction and cellular dynamics in diverse functional contexts, allowing to mimic complex and localized stimuli exerted by a dynamic extracellular microenvironment. By scaling the dimensions of the active magnetic pillars, our method can be applied to several biological studies, both on single cells and tissues, enabling a better understanding of the coupling between local external forces and intracellular biochemical pathways regulating cellular functions.

**Results**

**Magnetic micropillars: working principle and characterization**

The concept of the magnetic actuation is illustrated in Figure 1a, with reference to a single square group of magnetic pillars acting on a cell cultured on top. When a uniform in-plane magnetic field ($H_e$) is applied along the side of the pillars square, the couples of adjacent pillars along the field direction experiment an attractive force, arising from the proximity of magnetic charges of opposite sign. On the other hand, those perpendicular to $H_e$ feel a repulsive force, produced by the closer proximity of magnetic charges of the same sign (see Fig.1a). If a rotating field is applied, a continuous bending of the pillars occurs, exerting on cells a periodic and biaxial mechanical stimulus, compressive and tensile along perpendicular directions. Figure 1b shows an electron microscopy image of the active substrate developed in this paper, consisting in a two-dimensional array of square groups of four polydimethylsiloxane (PDMS) micropillars. The magnetic head is made of a 150 nm thick *Fe* film, sandwiched by two 50 nm thick $SiO_2$ layers (see Fig.1c). Each pillar is 10 μm high and 5 μm wide, with a minimum distance between adjacent pillars of 2 μm, in absence of magnetic field. The distance between neighboring groups is 6 μm, so as to neglect the magnetic interaction between them.

To estimate the magnitude of the magnetic forces, let us first consider the basic unit cell of each structure, i.e. a couple of adjacent pillars (see Fig.1d). Under the application of an in-plane field **$H_e$**, two adjacent pillars reciprocally interact, as the magnetization **M** in the first *Fe*-disk produces a magnetic stray field gradient on the second and vice-versa. For $\mu_0 H_e$ ranging between 10 and 100 mT, *Fe*-disks behave as magnetic dipoles in a single domain configuration (see Fig.1d), with **M** aligned to **$H_e$**, as resulting from micromagnetic simulations carried out with the software OOMMF (see *Methods*). The calculated stray field from the disks has been then used to evaluate the magnetic force **$F_M$** (see *Methods*). According to the physics of dipole-dipole interaction, **$F_M$** is attractive (positive) when **$H_e$** is directed along the line connecting the two pillars centers, while **$F_M$** is repulsive (negative) if **$H_e$** is perpendicular to that direction. For an external field $\mu_0 H_e = 50$ mT (the same used in biological experiments), we found an attractive force $F_M = 47.8$ nN when the field is oriented along the *x*-axis (see Fig.1d), and a repulsive force $F_M = -13.7$ nN for **$H_e$** applied along the *y*-axis. When **$H_e$** is directed at 45 degrees, a lower attractive force $F_M = 11.3$ nN is exerted. Note that the maximum attracting force is more than three times larger than the repulsive one, indicating that stimuli applied on cells are mainly compressive, similar to a "pinching" on a nanometric scale.

As expected, $F_M$ increases with $H_e$ because it favors the alignment of **M** along the field direction, resulting in larger magnetic moment of the *Fe*-disks. In Figure 1g, we report the simulated *x*-component of **$F_M$** (blue line) for a couple of pillars, when a $\mu_0 H_e$ field up to 100 mT is applied along the *x*-axis (see Fig.1d). The graph displays that $F_M$ increases with the external field, up to 73 nN. This demonstrates the possibility to tune the strength of the force with **$H_e$**, thus controlling the entity of pillars bending and, consequently, the mechanical nano-pinching applied on cells.

We measured the effect of **$F_M$** on the pillars deflection by optical microscopy, for variable **$H_e$** applied along the horizontal *x*-direction (see Fig.1e,f). The deflection ($\Delta x$) is measured as the difference between the distance separating the *Fe*-disks centers without field ($x_0$) and for applied $\mu_0 H_e$ ranging between 25-100 mT ($x_1$). The centers position is estimated by circular 2D fitting of the *Fe*-disks edges. As shown in Figure 1g (black line), $\Delta x$ increases with **$H_e$** up to 620±130 nm for $\mu_0 H_e = \pm 100$ mT, independently on the field polarity, as expected from the system symmetry. Moreover, the distance between disks centers when **$H_e$** is restored to zero ($x_0$), does not depend on the sequence of applied fields, in agreement with the negligible remanent magnetization of *Fe*-disks, causing a residual attracting force of just 1.2 nN. This is crucial for the reproducibility of the mechanical stimuli applied on cell, as both the initial condition and the applied stress are well defined and controlled univocally by the external field.

In order to check the consistency between pillars deflections and the strength of simulated forces, we model each pillar as a homogeneous cylinder[32]. The deflection is proportional to the force applied to the top of the pillars, i.e. $\Delta x = 2F_M/k$, where $k = (3/64)\pi E D^4/H^3$ is the elastic constant of the pillar, *D* and *H* are pillars diameter and height, respectively, and *E* is the Young modulus. The best fit of the experimental deflections (see red dashed-line in Fig.1g) is obtained for $E = 2.56$ MPa, slightly larger than the PDMS Young modulus (1.84 MPa)[32], but coherent with the presence of the *Fe*-disks deposited on top and some *Fe* on the side walls, which confers a larger rigidity to the pillars.

**Multiple mechanical pinching on individual cells**

To test the platform, NIH3T3 fibroblasts are cultured on the micropillars and exposed to mechanical nano-pinching by applying a rotating $\mathbf{H_e}$. Cells spread on few (3-5) 4-pillars groups (see Fig.2c), thus experiencing multiple stimulations at different points of the plasma membrane. At variance with passive microposts used so far [22,23,24], cells do not significantly alter the pillars' position due to the reduced aspect ratio (see Fig.1b). Biocompatibility has been assessed on cells cultured on the active substrate for three days. No evident change in viability and proliferation was observed.

During all the experiments a rotating field with amplitude $\mu_0 H_e$= 50 mT was applied. The rotation frequency ($f_F$) was 0.05 Hz, while the pinching frequency ($f_P$) was $2 \cdot f_F$ = 0.1 Hz, as the stress field is unchanged upon 180 degrees rotation of $\mathbf{H_e}$ (see frames 1 and 5 in Fig.2a). In order to properly visualize the mechanical nano-pinching, pillars were coated with fluorescent Cy5-fibronectin. Figure 2a shows the frames from a video, illustrating the different configurations of a group of 4-pillars, during the mechanical stimulation of a fibroblast. When $\mathbf{H_e}$ rotates, depending on the field orientation, a time-varying stress field is exerted on the cell, as pictorially depicted in Figure 2b. In frame 1, for $\mathbf{H_e}$ directed along *x*, a compressive (tensile) stress is applied along *x* (*y*); the pillars define a rectangle (dashed line in Fig.2a) stretched along *y*. Rotating $\mathbf{H_e}$ counterclockwise, at 45 degrees with respect to *x*, a weaker compressive stress, both along *x* and *y*, is applied (frame 2). Again, a biaxial stress like that of frame 1, but rotated by 90 degrees, is obtained for $\mathbf{H_e}$ at 90 degrees with respect to *x* (frame 3). Further rotating the field at 135 and 180 degrees, configurations of frames 4 and 5 are produced, which by symmetry are the very same of frames 2 and 1, respectively.

To highlight the dependence of the force on the field direction, the simulated magnetic force acting on each pillar is plotted (see Fig.2d) as function of the angle between the field and the x-axis ($\boldsymbol{\phi}$). For a rotating applied field with amplitude $\mu_0 H_e$= 50 mT, the force components ($F_x$ and $F_y$) are periodic in $\boldsymbol{\phi}$ and the simulated values are well fitted by sine functions (see Fig.2d). As expected, the maximum of the *x*-component ($F_x$ = 47.8 nN) is found for $\boldsymbol{\phi}$= 0 degrees, while at 90 degrees the force is less intense but repulsive ($F_x$ = -13.7 nN). By symmetry, $F_y$ is equivalent to $F_x$, but shifted by 90 degrees. Note that, the force acting on a single pillar is calculated (see *Methods*) just by considering the interaction between first neighboring pillars and neglecting the one along the diagonal of the square, which results in a negligible contribution (lower than 2 nN).

In turns, the maximum strain applied to the cell, assuming that the cell membrane locally follows the displacement of the disks as in Figure 2a, is about 5%.

**Nano-pinching affects nuclear morphology**

As shown in previous works[33,34], the application of forces and the alteration of substrates stiffness can affect the shape of the cell nucleus. For this reason, we first investigated the impact of mechanical nano-pinching by magnetic pillars on the nuclear morphology. As quantitative indicator of the cell nucleus morphology, we use the eccentricity ($\varepsilon$) of the nuclei projected area, extracted from fluorescence H2B-EGFP images (see *Methods*). H2B is a core histone, a nuclear protein responsible for the chromatin structure. The cells are imaged for 3 minutes before application of the mechanical stimulus, under a static $H_e$ at 45 degrees, producing a weak compressive stress both along *x* and *y* (see frame 2 in Fig.2a). Subsequently, cells undergo pillars stimulations and are imaged for 9 minutes with an acquisition rate of 0.5 frames per second (fps). Frames reported in Figure 3a show that the nucleus appears less elongated under mechanical stimulation ($t$ = 8 min) than before ($t$ = 2 min) and the corresponding eccentricity as function of time is plotted in Figure 3b. The average eccentricity before pinching ($\varepsilon_{BP}$), from 0 to 3 min, is 0.76, while during pinching it decreases to $\varepsilon_{DP}$= 0.72. $\varepsilon_{DP}$ is calculated as the average eccentricity between 9 and 12 min, in order to discard the transitory when the cell is adapting to the dynamic substrate. Remarkably, the same decrease of $\varepsilon$ is observed in all the 10 cells studied. Although the initial value of the eccentricity for each nucleus is different, the relative variation of $\varepsilon$, averaged over the 10 cases investigated, turns out to be $\Delta\varepsilon_R$ = -4.5±1 % (see Fig.3c). This suggests a sizable reduction of nuclear tension in response to periodic mechanical stimuli.

Nuclear transitions to a less elongated state happens with intrinsic dynamics, characterized by a transition time ($t_R$) required to the nucleus to adapt its shape, moving from a first "quasi" stationary eccentricity before pinching to a second "quasi" stationary value of $\varepsilon$ during pinching (see Fig.3b). The transition time measured for 10 cells is reported in Figure 3d. The average value of $t_R$ is 3.1 min, definitely much longer than the pinching period ($T_P$ = 10 s). These findings indicate that the nuclear response is not directly and elastically coupled to the mechanical stimuli applied to the cell membrane. Transmission appears mediated by active and slower cellular processes, as detailed in the next sections.

**Nano-mechanical stimulation affects nuclear and chromatin dynamics**

Alterations of nuclear shape, such as those presented above, are related to modifications in the nucleus-cytoskeleton coupling[35,36] which can also induce changes in nuclear motility and deformability. In order to elucidate this aspect, we investigated the effect of a periodic nano-pinching on nuclear plasticity by monitoring the nuclear area fluctuations, according to a procedure recently developed by some of the authors[26]. Individual H2B-EGFP positive cells were imaged (at 3 frames per minute) for 30 min before and during mechanical pinching.

From each video, we extracted the percentage nuclear area fluctuations (PNAF) vs. time, defined as the fluctuations from the mean value of the nuclei projected area (see *Methods*), which provide information on the nucleus plasticity and deformability. In Figure 4a the PNAFs for a single cell are reported, where a clear enhancement of the nuclear fluctuations during pinching is observed. Figure 4b shows the statistical distribution of PNAFs, measured on 10 different cells, before (black) and during (red) pinching. Remarkably, the distribution of PNAFs during mechanical stimulation is much broader than that before, with a standard deviation ($\sigma_{DP}$= 1.3%) which is more than two times larger than before pinching ($\sigma_{BP}$= 0.5%). The analysis of PNAFs demonstrates that cellular nucleus deformability increases during stimulation, indicating a reduction of nuclear pre-stress.

Beside the dynamics of nuclear morphology, we studied the impact of the nano-pinching on protein dynamics inside the nucleus. To this scope, we performed FRAP (Fluorescence recovery after photobleaching) on H2B-EGFP positive cells. The procedure (see *Methods*) consists in measuring the fluorescence recovery in a certain area of the nucleus (a circle with a diameter of 4 µm), after photobleaching of that region with high laser intensity. In Figure 4c, we compare the normalized intensity in the bleached area vs time, in cells subjected/not-subjected to mechanical pinching, bottom and top panels, respectively. The recovery fraction (see Fig.4d), calculated on 10 different nuclei, is faster during pinching, especially immediately after bleaching. This indicates a higher diffusivity and enhanced dynamics of H2B histone inside the nucleus during stimulation, thus suggesting that nano-pinching leads to a relatively more decondensed chromatin structure.

**Nano-pinching induces Actin reorganization and MKL cofactor translocation**

To investigate the role of active cellular processes responsible for the transmission of mechanical stimuli from the cell membrane to the nucleus, we first studied the effect of such stimuli on actin, one of the most abundant proteins in the cytoskeleton[37]. Figure 5a represents a cell transfected with RFP-Lifeact and imaged for 20 minutes before and after the activation of nano-pinching. The images show sizable variations of the cell morphology during stimulation, while reduced dynamics are observed before pinching. To put this finding on a quantitative basis, we performed RFP-Lifeact images correlation analysis (see *Methods*), both before and during pinching. It involves a pixel-by-pixel correlation of RFP-Lifeact maps, taken at 3 frames per minute, with the initial time frame (at $t=0$ s for data before pinching and at $t=20$ s for data during pinching). Then, the correlation coefficient is calculated for each cell as a function of time. In Figure 5b we compare the average correlation coefficient from 10 different cells, before (black curve) and after (red curve) the application of mechanical pinching. By performing a linear fit of the two curves, it is possible to quantify the faster decay of the correlation coefficient during stimulation, with a slope (see the inset in Fig.5b) 2.4 times higher than in the static case. A clear enhancement in actin images de-correlation is therefore observed during pinching. This demonstrates that the periodic mechanical stimuli exerted by the pillars affects actin dynamics, confirming that cytoskeleton plays a major role on the mechanic signal transmission towards the nucleus, not via direct mechanical coupling but through the reorganization of actin upon mechanical stimulation.

Finally, we investigated the effect of the mechanical stimuli on *megakaryoblastic acute leukemia factor-1* (MKL) translocation. MKL is a transcription cofactor, located in both the nucleus and cytoplasm, which can shuttle between the two in response to mechanical stimuli, thus bringing about alterations in gene transcription. Recent studies[16,38,39] have demonstrated that the actin configuration is related to MKL translocation. MKL moves to the cytoplasm when actin fibers are depolymerized while a translocation to the nucleus occurs when actin polymerizes into fibers. During the experiments, we imaged cells before the mechanical stimulation and after 30 min from the nano-pinching application. The images in Figure 5c show that MKL cofactor shuttles outside the nucleus in response to the mechanical stimuli. To better visualize the MKL translocation we subtracted the intensity map of MKL before pinching from that during pinching ($I_{DP}-I_{BP}$) and this difference is shown in Figure 5d. Moreover, we report the average MKL intensity ratio $I_{DP}/I_{BP}$ during and before pinching, both inside and outside the nucleus, from data acquired on 10 different cells (see Fig.5e). Data show an increase of the intensity ratio in the cytoplasm ($I_{DP}/I_{BP}=1.19$) and a decrease in the nucleus ($I_{DP}/I_{BP}=0.83$) when cells are pinched, thus demonstrating that MKL translocates out of the nucleus. This mechanism provides a possible path for mechanical stimuli to induce changes in gene expression by regulating shuttling of transcription factors/co-factors.

Remarkably, MKL translocation to the cytoplasm suggests actin depolymerization in response to cell pinching. This is in agreement with our findings on nucleus morphology and dynamics, as a less elongated and more dynamic nucleus reflects a reduction of the mechanical stress induced by the cytoskeleton, as expected by a depolymerized actin configuration[26].

**Conclusions**

In this paper we present a novel platform for in-vitro application of nano-mechanical stimuli with highly tunable spatio-temporal behavior at different points of the cell membrane. It consists in an active substrate for cell culture, made of groups of PDMS pillars with magnetic heads, whose deflection can be controlled at the nanometer scale by external magnetic fields. A rotating magnetic field produces a periodic biaxial strain field, corresponding to a nano-pinching of the cell at the focal adhesion points. For platform validation, we have studied the NIH3T3 cell response at a fixed pinching frequency of 0.1 Hz, corresponding to a maximum strain of 5% on cells. Our study reveals that nano-pinching induces changes in nuclear morphology, deformability and H2B core histone dynamics. Remarkably, the nuclear response to external forces does not result from a direct coupling between the cell membrane and the nucleus, but involves active cellular processes, such as actin reorganization and MKL cofactor translocation. In perspective, by proper scale up or down of the size of magnetic pillars and temporal modulation of the external magnetic field, our method can be exploited in a large variety of biological studies, from single cells to tissues, where the application of controlled and localized forces is required.

**Experimental Methods**

**Magnetic pillars fabrication**
The active substrate is made of Polydimethylsiloxane (PDMS) pillars with *Fe* ferromagnetic heads, patterned in repeated groups of four (see Fig.1b). Each pillar is 10 μm high, 5 μm in diameter, with a minimum distance of 2 μm (when no external magnetic field is applied). They are fabricated by replica molding from a *Si* substrate, patterned by photolithography and reactive ion etching. PDMS is cast on top of the mold and thermally cured at 80 degrees for 3 hours, before the peeling procedure. On top of PDMS a tri-layer of *SiO$_2$* (50 nm) / *Fe* (150 nm) / *SiO$_2$* (50 nm) is deposited by e-beam evaporation. The first *SiO$_2$* layer favors the adhesion of *Fe* on top of PDMS, while the second layer isolates the magnetic material from the biological environment. *Fe* is chosen as ferromagnetic material for pillars actuation due to its reduced toxicity together with a large saturation magnetization ($M_S$= 1.72·10$^6$ Am$^{-1}$).

**Micromagnetic simulations and Magnetic force calculation**
Simulations to quantify the force between the magnetic pillars are performed using OOMMF (Object Oriented Micro Magnetic Framework)[40]. The micromagnetic configuration of *Fe*-disks and the related magnetic stray field are calculated using standard parameters for *Fe*: saturation magnetization $M_s$= 1.72·10$^6$ A·m$^{-1}$, exchange stiffness A= 2.1·10$^{11}$ J·m$^{-1}$, damping coeffcient τ= 0.01 and null magneto-crystalline anisotropy.
A 20x20x20 nm$^3$ unit cell has been used. Although the exchange length of iron is 2.4 nm, this represents a reasonable compromise ensuring reduced computational times. We have checked that using cubic unit cells with a side length of 5 nm does not introduce major modifications in the simulated stray field.

The magnetic force is calculated from the stray field produced by the adjacent disks, according to the following equation[41]:

$$\mathbf{F}_M = \mu_0 \nabla(\mathbf{m} \bullet \mathbf{H}) = \mu_0 \int_V \nabla(M \bullet \mathbf{H}) dV \qquad (1)$$

where **m** is the magnetic moment of *Fe*-disks and **M** the magnetization, considered uniform all over the disk volume, according micromagnetic simulations (see Fig.1d). **H** is the total field (**H**= $\mathbf{H_d}$ + $\mathbf{H_e}$), calculated using OOMMF, resulting from the sum of the stray field generated by the adjacent pillars ($\mathbf{H_d}$) and the external magnetic field ($\mathbf{H_e}$). The integration is performed numerically (with a custom written code in *MATLAB*) over the *Fe*-disk volume (V).

**Cell Culture and Plasmid Transfections**

NIH3T3 fibroblasts stably expressing H2B-EGFP were cultured in low-glucose Dulbecco's Modified Eagle Medium (Gibco; LifeTechnologies) supplemented with 10% (vol/vol) FBS (Gibco; Life Technologies) and 1% penicillin-streptomycin (Gibco; Life Technologies) at 37 °C and 5% $CO_2$ in humid conditions. Cells were transfected with RFP-Lifeact or mcherry-MKL by electroporation (Gibco; Life Technologies), the day before the experiment. Cells were trypsinized (Gibco; Life Technologies) and seeded on micropillars coated with 20 μg/ml of Bovine Serum Albinum (BSA, Sigma Aldrich) and Fibronectin (Gibco; Life Technologies) for 3 h followed by 100 μg/ml of fibronectin for 1 h. Before imaging, the chip was inverted in a petri dish, on two *parafilm* spacers to avoid contact between the cells and the bottom of the dish. A special $CO_2$-independent medium (Gibco; Life Technologies, catalogue number: 18045) was used during the experiments.

**Imaging, Magnetic field application and Image Processing**

The dish containing our active substrate with the cells cultured on the pillars was placed under a NikonA1R Confocal microscope with 20x and 40x objectives. The rotating magnetic field is applied with two $Nd_2Fe_{14}B$ permanents magnets, mounted on a rotating 3D-printed support, which was mechanically isolated from both the microscope and the sample stage. The Magnetic field rotation is provided by a stepper motor and regulated by an *Arduino UNO* microcontroller. Acquisition is performed in bright field and confocal mode with different acquisition rates according to the experiments: the fast dynamics are imaged with a rate of 0.5 fps, while the slow dynamics at 3 frames per minute. The *z*-depth for confocal imaging (Nucleus morphology, Nuclear area fluctuations, Images correlation and MKL-signal imaging) is ~500 nm. A custom written code in *MATLAB* was used for H2B-EGFP image thresholding, projected nuclear area calculation, geometrical parameters extrapolation and image correlation analysis. mcherry-MKL intensity and FRAP analysis were performed with the *ImageJ* software. MKL-intensity images subtraction was performed using *MATLAB*. The pillars deflections in Figure 1e,f were measured with NIKON-eclipse optical microscope equipped with a 60x immersion objective.

**Nuclear Area fluctuations analysis**

Absolute projected nuclear area was first measured by thresholding the average intensity projection of confocal *z*-slices of the H2B-EGFP positive nuclei. This projected area was then plotted as a function of time and fitted with third-order polynomial curves in ORIGIN. The residual values were divided by the value of the polynomial at each time point and multiplied by 100 to obtain the percentage nuclear area fluctuations (PNAF). Such PNAFs from n=10 cells and all the time points, were combined to obtain a normal distribution, either from data acquired before and during pinching. Standard deviations of the PNAFs distributions before and during pinching ($\sigma_{BP}$ and $\sigma_{DP}$) indicate the amplitude (in percentage) of area fluctuations in the two cases.

**Images Correlation Analysis**

A pixel-by-pixel images correlation analysis was performed to investigate RFP-Lifeact dynamics (see Fig.5b). Starting from a reference frame, we acquired images for 20 minutes, at 3 frames per minute, and a 2D correlation coefficient between each frame and the reference one is calculated, according to the following equation:

$$c = \frac{\sum_m \sum_n (A_{mn} - \bar{A})(B_{mn} - \bar{B})}{\sqrt{\left(\sum_m \sum_n (A_{mn} - \bar{A})^2\right)\left(\sum_m \sum_n (B_{mn} - \bar{B})^2\right)}} \qquad (2)$$

where $c$ is the 2D correlation coefficient, while $A_{mn}$ and $B_{mn}$ are the matrix elements representing the pixels of the two images. A is the image of the reference frame, while B is the image taken at different times during acquisition. $\bar{A}$ and $\bar{B}$ are the average intensity of the two images. The subtraction of the average value reduces the impact of photobleaching on the estimate of the images correlation.

In this way, we calculated for each cell the correlation curve, i.e. the evolution of $c$ vs. time (see Fig.5b).

**FRAP Analysis**

A fluorescence recovery after photobleaching (FRAP) experiment was performed on H2B-EGFP labeled cells. First, a circular region (~4 µm in diameter) in the nucleus was photobleached. Then images were acquired at 12 frames per minute during the first 5 minutes, to capture the fast dynamics of the fluorescence recovery, and then at 3 frames per minute for 20 minutes. Using the *ImageJ* software, the fluorescence intensity in the photobleached region is computed at each time frame, before and after photobleaching. A normalized intensity ($I_{norm}$) is calculated using the following formula:

$$I_{norm}(t) = \frac{I(t) - I_B}{I_{pre-bleach} - I_B} \bullet \frac{T_{pre-bleach} - I_B}{T(t) - I_B} \qquad (3)$$

where $I(t)$ is the measured intensity in the bleached area, $I_B$ is the background intensity, $I_{pre-bleach}$ is the average intensity before photobleaching in the bleached region, $T_{pre-bleach}$ the intensity of the whole nucleus before bleaching and $T(t)$ the total intensities of the whole nucleus as function of time. The first factor in the equation allows to calculate the recovery fraction, normalizing $I(t)$ to the initial value and rescaling it between 0 and 1. The second factor, instead, allows to compensate the general tendency to underestimate the fluorescence recovery in FRAP experiments, due the overall bleaching of the cell, by normalizing the intensity in the bleached area also to the average intensity from the nucleus.

**Acknowledgments**

We thank E. Albisetti, P.P. Sharma, D.V. Conca and A. Ravasio for the fruitful discussions. M.M. thanks G. D'Abrosca for her skilful graphic support. M.M., D.S.J. and G.V.S. thank "MBI microscopy core" for the optical components they provided and "MBI wet lab core" for actively maintaining the cell culture room that was used. M.M., D.P. and R.B. thank C. Somaschini and G. Iseni for the technical support. This work was funded by MIUR via the project "ATR and ATM-mediated control of chromosome integrity and cell plasticity" (Project No. 2015SJLMB9) and by Fondazione Cariplo via the project UMANA (Project No. 2013-0735). The fabrication of the devices was performed at PoliFab, the micro- and nanofabrication facility of "Politecnico di Milano".

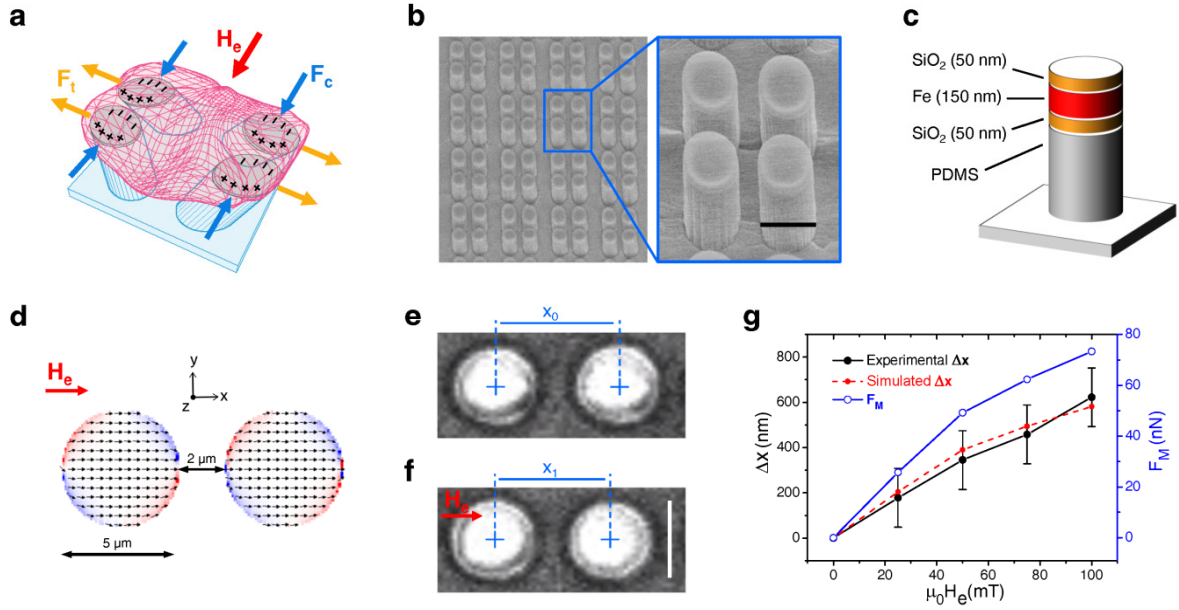

**Figure 1. Magnetic pillars working principle and characterization. a** Sketch of the device showing a group of 4 *Fe*-coated pillars with a cell cultured on top. The application of a uniform external magnetic field ($H_e$) induces adjacent pillars interaction, producing pillars bending and thus applying mechanical stimuli on cells. Compression of adjacent pillars occurs in the field direction, while they are stretched in the perpendicular one. On top of *Fe*-disks, the magnetic charges induced by $H_e$ are depicted. Blue and orange arrows represent the magnetic force ($F_M$) components (compressive and tensile respectively) exerted on each pillar. **b** Scanning electron microscopy images of the device. *Fe*-coated PDMS micropillars are 10 μm high, 5 μm in diameter and spaced 2 μm (minimum distance, when $\mu_0 H_e$= 0 mT). On the right: zoom on a single group of 4-pillars. **c** Sketch of a single magnetic pillar with a trilayer of *SiO₂* (50 nm) / *Fe* (150 nm) / *SiO₂* (50 nm) deposited on top. **d** Micromagnetic configurations (simulated using OOMMF) of two adjacent *Fe*-disks on top of PDMS pillars, when an external magnetic field ($\mu_0 H_e$= 50 mT) is applied along the *x*-axis. The arrows represent the local magnetization direction, while the red-white-blue scale color refers to the *y*-component of the magnetization. **e-f** Optical microscopy images of two adjacent pillars, comparing the distance between centers without ($x_0$, panel **e**) and with ($x_1$, panel **f**) the application of $\mu_0 H_e$= 50 mT along the *x*-axis. **g** On the left *y*-axis: experimental (black line) and simulated (red dashed-line) deflection ($\Delta x = x_0 - x_1$) of magnetic pillars as function of the external field ($H_e$), directed along the line connecting adjacent *Fe*-disks centers (*x*-axis in Fig.1d). The relatively large uncertainty of experimental data arises from the limited resolution of optical microscopy, combined with shape defects of the pillars, which hinder the 2D fitting for the determination of the center position. On the right *y*-axis: calculation of the *x*-component of the magnetic force ($F_M$, blue) between two adjacent magnetic pillars as function of $H_e$. Scale bars: 5 μm (**b**, **f**).

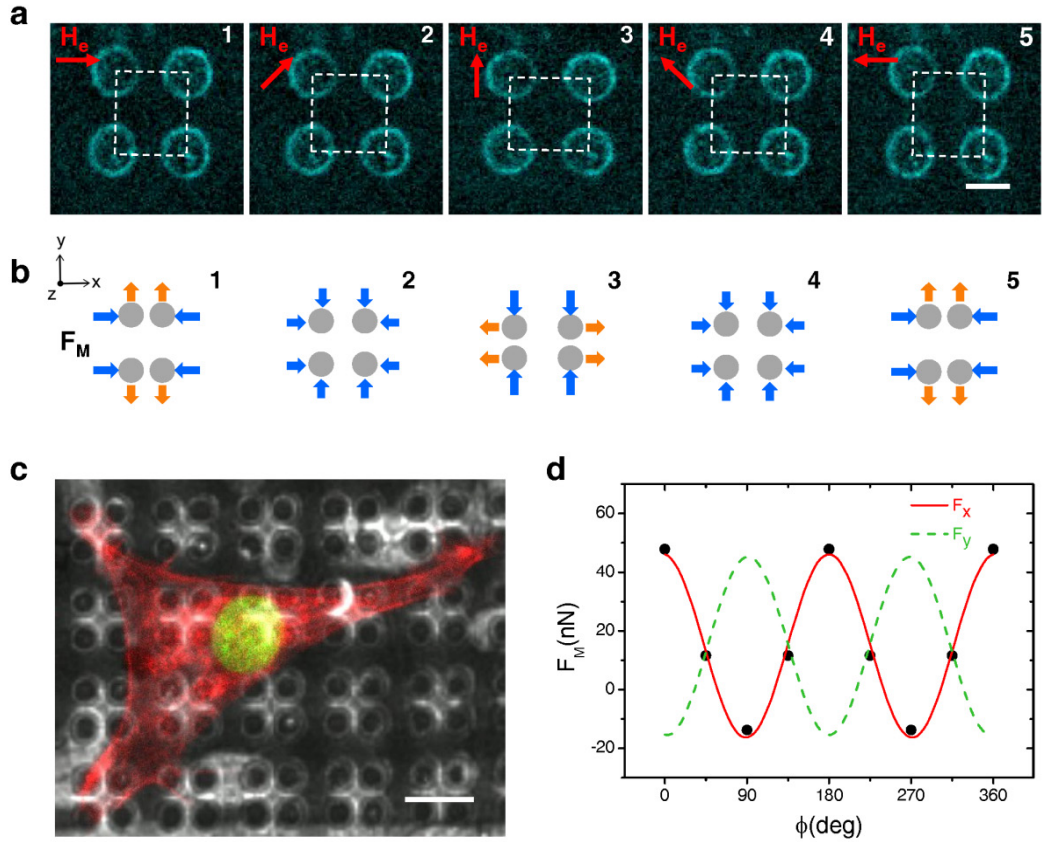

**Figure 2. Magnetic pillars exert nano-pinching on NIH3T3 cells. a** Frames from a video showing different configurations of a square group of magnetic pillars with a fibroblast cell plated on top, when a rotating $\mu_0 H_e$ = 50 mT is applied. Pillars, coated with Cy5-fluorescent fibronectin, sequentially attract and repel in vertical and horizontal directions according to the orientation of $\mathbf{H_e}$. **b** Scheme of the forces exerted on a group of four pillars. The blue (and orange) arrows represent the direction of the attractive (and repulsive) force components exerted on the cell by each pillar. **c** Optical image showing the device with the magnetic *Fe*-coated pillars and a single NIH3T3 cell transfected with RFP-Lifeact (red fluorescence) and H2B-EGFP (green fluorescence). **d** Simulations of the magnetic force ($\mathbf{F_M}$) exerted on magnetic pillars, as function of the field direction ($\mathbf{H_e}$), when a rotating $\mu_0 H_e$ = 50 mT is applied. The simulated data (black dots) are fitted with a sinusoidal curve for the *x*-component ($F_x$, red-line). By symmetry, the *y*-component ($F_y$, dashed green-line) has the same behavior of $F_y$, but displays a phase shift of 90 degrees. Scale bars: 5 μm (**a**), 20 μm (**c**).

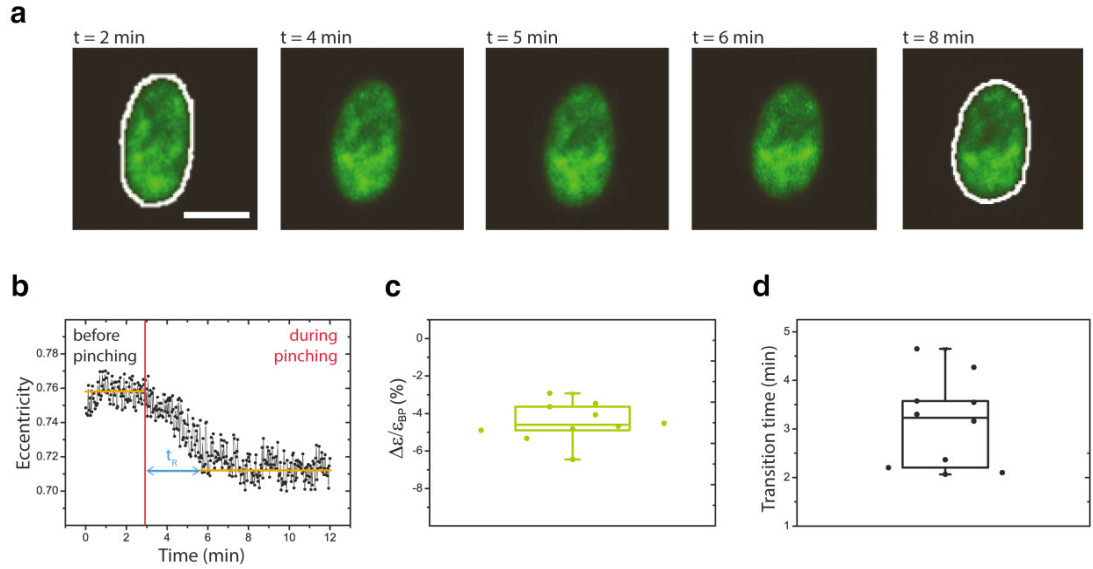

**Figure 3. Mechanical pinching exerted by magnetic pillars affects nucleus morphology. a** Frames from a video showing a NIH3T3 cell nucleus (H2B-EGFP green fluorescence), during an experiment. The application of a rotating field ($\mu_0 H_e$= 50 mT) at *t* = 3 min affects the nucleus morphology, which becomes less elongated. Cells are imaged for 3 minutes before pinching and for 9 minutes during pinching. The white lines identify the nucleus profile. **b** Nucleus projected area eccentricity as function of time, before (*t*= 0-3 min) and during (*t*= 3-12 min) mechanical pinching. $t_R$ is the response time of the nucleus to a less elongated state. The orange lines represent the average eccentricity before and during pinching. **c** Box plot for the percentage nucleus projected area eccentricity variation $\Delta\varepsilon_R = \frac{\varepsilon_{DP} - \varepsilon_{BP}}{\varepsilon_{BP}}$, where $\varepsilon_{BP}$ and $\varepsilon_{DP}$ are respectively the average eccentricity before pinching (*t*= 0-3 min, see Fig.3b) and during pinching (*t*= 9-12 min). $\Delta\varepsilon_R$ is calculated for a batch of n= 10 cells (data acquired in three different experiments). The bottom and top of the box represent the first and third quartiles, whereas the line inside the box represents the median. The ends of the whiskers correspond to the lowest/highest data point of the distribution. **d** Box plot for the transition time ($t_R$) of the nucleus projected area to a lower eccentricity "quasi stationary state" (see Fig.3b), extrapolated by a batch of n= 10 cells. The bottom and top of the box represent the first and third quartiles, whereas the line inside the box represents the median. The ends of the whiskers correspond to the lowest/highest data point of the distribution. Scale bar: 5 μm.

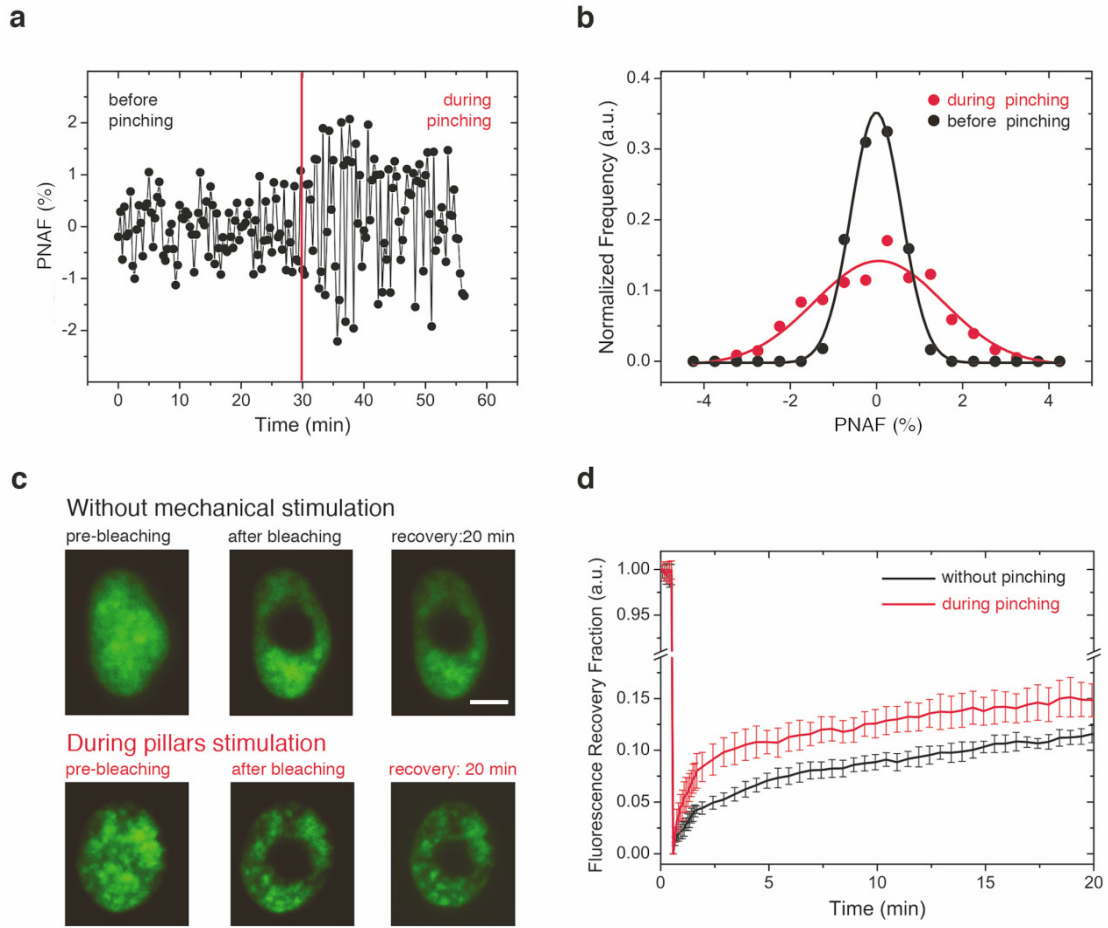

**Figure 4. Magnetic pillars stimulation affects nuclear area fluctuations and H2B dynamics. a** Percentage nuclear area fluctuations **(**PNAFs) vs time of a cell before and during pinching. The red line represents the time at which rotation of $\mu_0 H_e$= 50 mT is turned on. **b** Black and red dots represent the distribution of combined percentage nuclear area fluctuations (PNAFs) for n= 10 cells and all the time points, before and during pinching respectively (data acquired in four different experiments). In order to disregard transitory effects, the statistics is related to data acquired from 10 to 30 min after the field rotation is turned on. Continuous lines are Gaussian fittings. **c** Frames showing H2B-EGFP fluorescence intensity upon photobleaching and recovery, without and with mechanical pinching of cells. The bleached ROI is a circle with a diameter of 4 µm. **d** Fluorescence recovery curves for nuclear H2B-EGFP signal (mean on n= 10 cells), without (black) and with (red) the application of a rotating $\mu_0 H_e$= 50 mT (data acquired in four different experiments). The error bars represent the standard deviations from the mean. Scale bar: 5 µm.

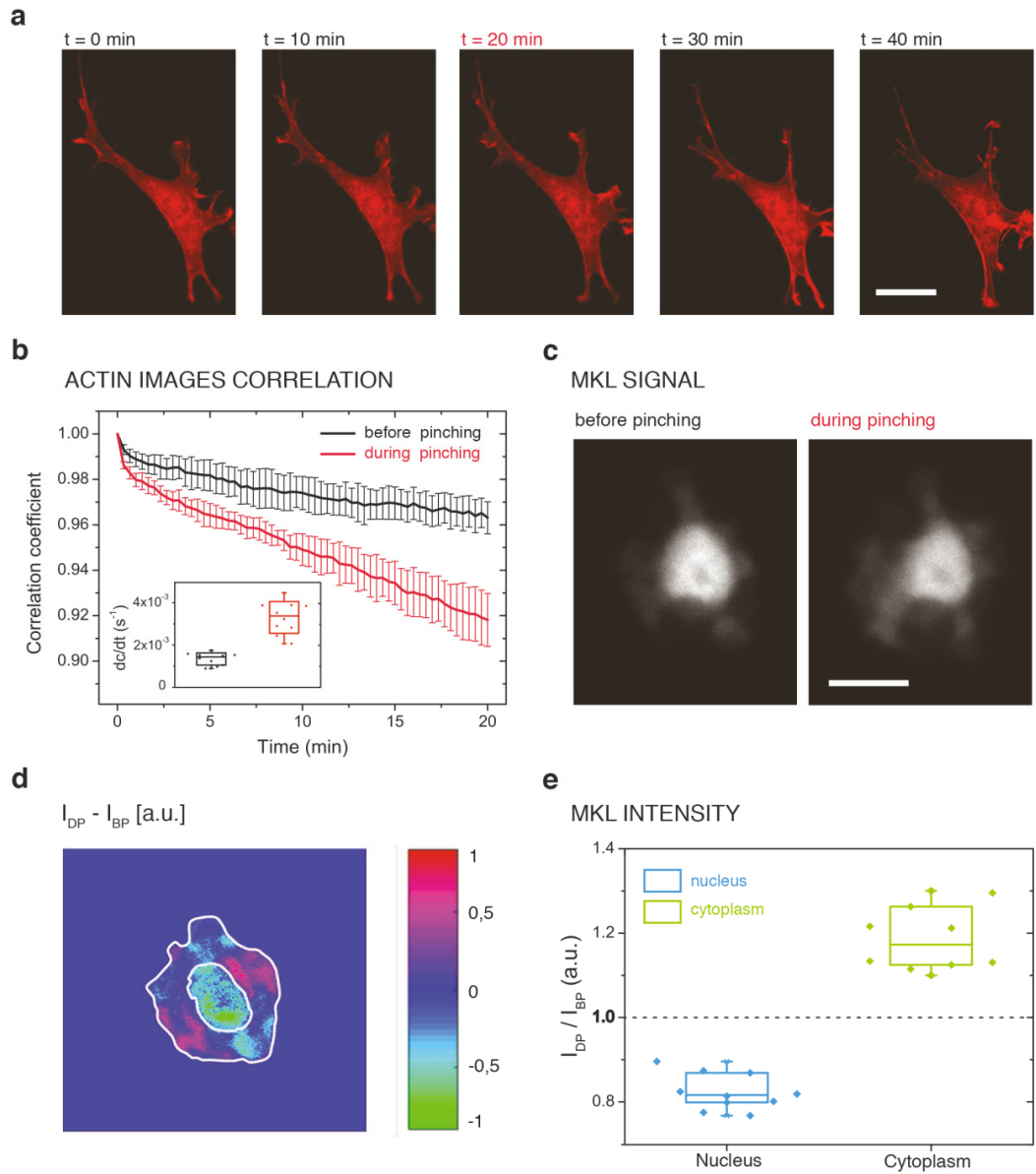

**Figure 5. Magnetic pillars stimulation induces actin reorganization and MKL translocation. a** Frames from a video showing a NIH3T3 cell (RFP-Lifeact red fluorescence), during an experiment. The cell is imaged for 40 minutes, before (0-20 min) and during (20-40 min) mechanical pinching. Upon application of a rotating field ($\mu_0 H_e$= 50 mT) at $t$= 20 min, faster actin dynamics are observed. **b** Images correlation vs time of RFP-Lifeact red fluorescent signal from the reference frame, before (black, reference frame at $t$= 0 min) and during (red, reference frame at $t$= 20 min) pinching, calculated for n= 10 cells (data acquired in four different experiments). The correlation coefficient is calculated according to Equation 2, performing a pixel-by-pixel analysis. The reference frames during pinching corresponds to the time point at which the field rotation ($\mu_0 H_e$= 50 mT) is turned on. Error bars represent the standard deviations from the mean. The inset shows box plots for the linear fitting of images correlation coefficient slopes (dc/dt), calculated for 10 cells before (black) and during (red) pinching. The bottom and top of the box represent the first and third quartiles, whereas the line inside the box represent the median. The ends of the whiskers correspond to the lowest/highest data point of the distribution. **c** Optical images showing mcherry-MKL signal in the cell nucleus and cytoskeleton, before and during the mechanical stimulation. The rotating field ($\mu_0 H_e$= 50 mT) is applied for 30 min before the acquisition of the second frame. **d** Color map of MKL signal, subtraction the intensity during ($I_{DP}$) and before ($I_{BP}$) pinching. **e** Box plots for the MKL relative intensity, calculated as the ratio between the intensity during and before pinching ($I_{DP}/I_{BP}$) for n= 10 cells, respectively in the nucleus (blue) and cytoplasm (green). The MKL intensity both inside the nucleus and in the cytoplasm is calculated as the average of 5 different circular ROIs with a diameter of 2 μm. The bottom and top of the box represent the first and third quartiles, whereas the line inside the box represents the median. The ends of the whiskers correspond to the lowest/highest data point of the distribution. Scale bar: 20 μm.

17